\documentclass[aps,prd,twocolumn,10pt,showpacs]{revtex4}
\usepackage[mathcal]{euscript}
\usepackage[latin1]{inputenc}
\usepackage{latexsym}
\usepackage{amsmath}
\usepackage{hyperref}
\usepackage{amsmath}    
\usepackage{graphicx}   
\usepackage{verbatim}   
\usepackage{color}      
\usepackage{subfigure}  
\usepackage{bm}
\usepackage{amssymb,overpic}
\usepackage{bbm}
\usepackage{cancel}

\begin{document}
\setcounter{topnumber}{1}

\title{Hybrid classical-quantum formulations ask for hybrid notions}

\author{Carlos Barcel\'o}
\affiliation{Instituto de Astrof\'{\i}sica de Andaluc\'{\i}a (IAA-CSIC), Glorieta de la Astronom\'{\i}a, 18008 Granada, Spain}
\author{Ra\'ul Carballo-Rubio}
\affiliation{Instituto de Astrof\'{\i}sica de Andaluc\'{\i}a (IAA-CSIC), Glorieta de la Astronom\'{\i}a, 18008 Granada, Spain}
\author{Luis J. Garay}
\affiliation{Departamento de F\'{\i}sica Te\'orica II, Universidad Complutense de Madrid, 28040 Madrid, Spain}
\affiliation{Instituto de Estructura de la Materia (IEM-CSIC), Serrano 121, 28006 Madrid, Spain}
\author{Ricardo G\'omez-Escalante}
\affiliation{Departamento de F\'{\i}sica Te\'orica II, Universidad Complutense de Madrid, 28040 Madrid, Spain}

\begin{abstract}

We reappraise some of the hybrid classical-quantum models proposed in the literature with the goal of retrieving some of their common characteristics. In particular, first, we analyze in detail the Peres-Terno argument regarding the inconsistency of hybrid quantizations of the Sudarshan type. We show that to accept such hybrid formalism entails  the necessity of dealing with  additional degrees of freedom beyond those in the straight complete quantization of the system. Second, we recover a  similar enlargement of degrees of freedom in the so-called statistical hybrid models. Finally,  we use Wigner's quantization of a simple model to illustrate how in hybrid systems the subsystems are never purely classical or quantum. A certain degree of quantumness (classicality) is being exchanged between the different sectors of the theory, which in this particular unphysical toy model makes them undistinguishable.

\end{abstract}
\maketitle


\section{Introduction}

After almost 100 years since  the formulation of quantum mechanics we still have to face unsettled issues, particularly concerning the elucidation of its regime of applicability.
 
On one extreme, classical mechanics is openly an effective but very convenient description of the macroscopic world. Under clear approximations one can treat certain systems as being composed of interacting classical points (e.g. the solar system) or as continuous classical entities (e.g., fluid mechanics). Based on these notions we have developed the concept of a curved spacetime as a classical and continuous arena in which everything else seems to take place.   

On the other extreme, quantum mechanics appears as a potentially fundamental description of the microworld and, adopting a reductionist view, also of the macroworld. However, arguably, quantum mechanics is not self-contained as it incorporates notions from the classical realm in its very formulation and its operational verification. Moreover, we are certain of the faithfulness of quantum mechanics as an accurate model of the world only in microphysical experiments monitored by perfectly classical apparatus. If only because it incorporates classical notions, one could expect quantum mechanics to be also an effective, but very convenient, nevertheless, description of the microworld.

We do not know how to establish a clear sharp border between the micro- and the macroworlds; it is difficult even to know when a variable is going to behave  classically or quantum mechanically.  
What is certain is that the cleanly explored limits are still far apart: e.g., in terms of interference of collection of atoms, one has to go from the observed interferences of beams of  molecules made of $10^3$ atoms to the clearly classical behavior of collections of $10^{23}$ atoms. Therefore, from a purely logical and observational point of view, it is still possible that there exists one, or several, depending on the situation, hybrid classical-quantum theory (not derivable from pure quantum mechanics) which provides a better description of the yet-to-come mesoscopic experiments. Furthermore, such hypothetical theory would surely be more apt to describe our perception of the world, in which its macro-variables seem  to have as much relevance in its evolution as its microvariables~(see e.g. the discussion in~\cite{Ellis2011}). In this paper, among other specific developments, we want to put forward the view that the study of hybrid classical-quantum constructions suggests that it is indeed possible to build consistent hybrid theories, but at the (reasonable) cost of allowing them to be populated by subsystems which are neither purely classical nor purely quantum: Every subsystem would have a certain degree of classicality and of quantumness, with pure notions being applicable only to ideal situations. Moreover, the theory would not be  a suitable hybrid limit of the straightforward fully quantum theory: It would exhibit new physical properties.

The possibility of formulating a consistent hybrid classical-quantum theory has been approached in many different ways. In all these approaches one starts with initially separated purely classical and quantum sectors and then makes them interact in order to analyze the outcome. Without pretending to be exhaustive, we can classify these approaches in the following categories: (1) approaches that try to maintain the use of quantum states (or density matrices) to describe the quantum sector and trajectories for the classical sector~\cite{BoucherTraschen1988,Anderson1995}, (2) those that first formulate the classical sector as a quantum theory~\cite{Koopman1931,Neumann1932,Sudarshan1976} and then work with a  formally  completely quantum system~\cite{Sherry1978,Sherry1979,Gautam1979,PeresTerno2001,Terno2006}, (3) conversely, those that first formulate the quantum sector as a classical theory~\cite{Heslot1985} and then work with a formally completely classical system~\cite{Elze2011,Alonso2010,Alonso2011,Alonso2012}, and (4) approaches that take the quantum and the classical sectors to a common language and then extend it to a single framework in the presence of interactions, for instance, using Hamilton-Jacobi statistical theory for the classical sector and Madelung representation for the quantum sector~\cite{HallReginatto2005,Hall2008,Chua2012} or
modeling classical and quantum dynamics starting from Ehrenfest equations~\cite{Bondar2012}. This classification  is not sharp and in some cases is subject to interpretation, but it may be useful as a way to organize the possible procedures and conceptual viewpoints in the enterprise of  constructing a hybrid theory.

Approaches of type 1 have been shown to lead to a number of inconsistencies~\cite{CaroSalcedo1998}. On the one hand, the standard semiclassical treatment~\cite{BoucherTraschen1988} does not transfer the primary fluctuations in the quantum sector into the classical sector. This is, for example, problematic for the inflationary paradigm of generation of structures in the universe~\cite{Brandenberger1984,Sudarsky2010}. On the other hand, more refined semiclassical treatments~\cite{Aleksandrov1981,BoucherTraschen1988} in which the primary quantum fluctuations do affect the classical sector also lead to inconsistencies. For instance, the dynamics is dictated by a bracket failing to satisfy Jacobi's identity and Leibniz's rule. Under reasonable hypotheses it has been shown that there is no Lie bracket for a semiquantized theory of this form~\cite{Salcedo1995,CaroSalcedo1998,Sahoo2004,Dass2009,Salcedo2012}. If one permits the bracket to be of non-Lie type, it can be shown that the theory will loose the positivity of its density matrix~\cite{BoucherTraschen1988,CaroSalcedo1998} (positive operators could have   non-positive expectation values). In~\cite{CaroSalcedo1998} it is also argued that induced fluctuations in the classical sector would spoil the commutativity of the classical variables. However, this is based on assuming that the total Hamiltonian is bounded from below; as we will see, in the Koopman-von Neumann-Sudarshan approach this boundedness is lost~\cite{PeresTerno2001}, so that one can accommodate commuting variables with induced fluctuations.

Approaches in category 2 deal with a perfectly defined quantum theory based on Koopman-von Neumann-Sudarshan~\cite{Koopman1931,Neumann1932,Sudarshan1976} translation of classical mechanics into the language of Hilbert spaces. Thus, previous criticisms based on the formulation of a well-defined bracket do not apply to it. However, within this approach, the work of Peres-Terno is often mentioned as an important objection to the consistency of hybrid configurations. They proved that, in the presence of interaction, Heisenberg's equations of motion of the canonical-variable operators (Heisenberg picture) are necessarily different from what would have been obtained by assuming the two sectors to be quantum. From this they conclude (as we will see, somewhat precipitately) that the classical (Ehrenfest) limit of the classical-quantum (CQ) system will be necessarily different from that obtained directly from the corresponding quantum-quantum (QQ) system. On the other hand, they argue that these two limits should coincide, that is, that any hybrid theory should comply with this {\em definitive benchmark}, as they call it. Then, they finally conclude that hybrid systems (at least of this kind) are unphysical.

Here, first of all, we want to stress that this benchmark only makes  sense if one tacitly assumes that a hybrid system should be just some suitable limit of a purely quantum system. If this is not assumed, we can invert the logic and see the failure to satisfy this condition as a clear sign of new physics in the CQ interplay. Second, as we will see in Sec.~\ref{Sec:Koopman}, the Peres-Terno model has many interesting features, being more complex than just exposed.  When looking at the quadratic Peres-Terno model in more detail, one realizes that {\em it can} be made to comply with the above benchmark by just adding some constraints to the system. In fact, to add constraints to the system is more than reasonable because its very formulation incorporates more degrees of freedom than the QQ theory with which it is compared. Nonetheless, in a second  twist, we will show that this hybrid system still exhibits some sort of refined Peres-Terno no-go result: One cannot require that the dynamical equations for the second-order momenta (i.e. the dispersions) be  equal to those in the QQ system (which for these system are equal to the CC ones). Going beyond quadratic interactions we also show that there is no  way in general to fulfill the Peres-Terno benchmark by adding constraints. Overall, we are led to conclude that either the evolution of the expectation values or the evolution of the dispersions will exhibit some new physics beyond both QQ and CC systems. 

Let us mention that in a series of papers Sudarshan first proposed this hybrid formalism as a way of understanding quantum measurement~\cite{Sherry1978,Sherry1979,Gautam1979}. In doing that he restricted the types of interaction terms that one can consistently introduce. The guiding principle for these restrictions was what he called the {\em principle of integrity} of the classical variables. In these analyses it is explicitly assumed that the resulting hybrid theory is more than the limit of a purely quantum theory.

Approaches of type 3, as formulated, for example, in~\cite{Elze2011}, 
avoid the problems raised in approaches of type 1, but mainly because of a shift of perspective. We have a classical Hamiltonian system for the variables $q,p,a_i$, and $a_j^*$, where $a_i$ and $a_j^*$ are the coefficients of an expansion of the quantum state in the Hilbert space and $q$ and $p$ are the genuinely classical variables. This formulation is, in fact, equivalent to the standard semiclassical treatment so that it does not transfer the quantum fluctuations to the classical variables.  When changing to a statistical description, the transfer of fluctuations to the classical sector is allowes. In Ref.~\cite{Elze2011} it was already noticed that the interaction makes the system migrate beyond the border marked by the very definition of the space of all possible unit-norm quantum states (see also~\cite{Buric2012} for a motivation of the form of these interaction terms). This is natural since the evolution of the quantum sector becomes, in general, nonunitary when interacting with external variables. In addition, for consistency, the meaning of observable has to be extended from what is directly suggested by the CQ system, i.e., a function on the classical phase space with values in self-adjoint operators in the quantum Hilbert space.  With this definition, if $A$ is an observable of the hybrid system, then its time derivative is not an observable in general.  

Moreover, Salcedo~\cite{Salcedo2012} has recently shown that the evolution of a statistical mixture of configurations would evolve differently for different representations of the initial density matrix of the quantum sector. This is so because a distribution function in configuration space has more information than a density matrix. If something of this sort were at work in nature, one would be able to identify special  sets of states in the Hilbert space  as the ones consistent with the observed evolution. One can interpret this special situation as a bad sign for the hybridization or, instead, as suggesting that the hybrid theory needs to be complemented with the selection of a natural basis  in the Hilbert space, something that resonates with the notion of ``pointer basis'' in decoherence theory.

Finally, approaches in category 4 are based on the well-known fact that both classical statistical mechanics and quantum mechanics can be cast in a fluid-mechanics form using a density and a nonrotational velocity flow as canonical field variables~\cite{HallReginatto2005,Hall2008}. Again, this formulation bypasses the bracket problem as the definition of observables goes beyond being a tensor product of observables in the quantum and classical sectors. In the presence of interaction the distinction of which is the classical and which is the quantum sector is blurred~\cite{Salcedo2012}. This approach has the same problem, or characteristic, as the previous one: A statistical mixture has much more information than a standard density matrix.

All in all, in our view the overall image that shows up from these analyses and the discussion in this paper is the following.
\begin{itemize}
\item
Whereas it is certainly possible to construct hybrid systems, these constructions typically ask  for the introduction of hybrid concepts absent in a straight classical-quantum product.
These hybrid theories are not derivable from a straightforward purely quantum theory: They incorporate new physics. This explicitly warns us about the toy-model nature and heuristic character of the different  frameworks analyzed above.
\item
The construction of a consistent hybrid formulation is far from unique, so it would strongly need feedback from mesoscopic experiments. This is not strange after all. The new parameters would specify how the (close to) quantum sector would back react on the (close to) classical sector. The (close to) classical sector would typically be  described by  (close to) macroscopic variables that would have to be specified with enough detail to see what their effects would be on the (close to) quantum system and vice versa.
\end{itemize}

The rest of the paper is organized as follows.  Section~\ref{Sec:Koopman} is devoted to making a detailed analysis of approach 2, which is very suggestive of the situation one faces when dealing with hybrid systems. In particular, we reappraise the argument raised by Peres and Terno regarding the inconsistency of constructing a hybrid theory within the Koopman-von Neumann-Sudarshan formalism. Then, Sec.~\ref{Sec:Others} reviews the statistical consistency problem raised recently by Salcedo.
By changing the point of view, we show that this problem can be seen instead as a defining generic characteristic of hybrid approaches. Section~\ref{Sec:Wigner} takes the most simple model of a hybrid system one can imagine and describes quantitatively how the classicality or quantumness is being interchanged between the two sectors. We finish with a summary and some final conclusions.

\section{Sudarshan hybrids \label{Sec:Koopman}}

The starting point of this particular hybrid scheme is the Koopman-von Neumann-Sudarshan formulation of classical mechanics~\cite{Koopman1931,Neumann1932,Sudarshan1976,Mauro2003}. This formulation translates the usual description of classical mechanics in terms of symplectic manifolds to a quantum-mechanical language by associating to each physical observable  a self-adjoint operator in a suitable Hilbert space and by implementing the time evolution as a unitary operator. Once the classical sector is treated formally as quantum, one can describe a CQ interaction by using the tensor product of Hilbert spaces, as in a pure quantum-mechanical theory~\cite{Ballentinebook}. It is in developing this program that one has to face a number of difficulties.

To make the discussion self-contained let us briefly summarize the 
Koopman-von Neumann-Sudarshan formalism. For the sake of simplicity we shall deal with one-dimensional systems although the discussion can be easily generalized to an arbitrary number of degrees of freedom. The quantum sector will be represented by a Hilbert space $\mathcal{H}_{\mbox{\tiny Q}}$ with the standard position and momentum operators,
\begin{equation}
[\hat{x},\hat{k}]=i\hbar~.
\end{equation}
In the classical sector we  will also have position and momentum operators  
defined over a Hilbert space $\mathcal{H}_{\mbox{\tiny C}}$, but in this case they commute,
\begin{equation}
[\hat{q},\hat{p}]=0~.
\label{eq:clasop}
\end{equation}
The elements of the pre-Hilbert space $\overline{\mathcal{H}}_{\mbox{\tiny C}}$ are  taken to be the ``classical'' wave functions $\psi(q,p)$, whose square equal the classical distribution functions:  
\begin{equation}
 |\psi|^2=\rho_{\mbox{\tiny C}}(q,p)~.
\label{eq:claswave}
\end{equation}%
The action of the classical operators (\ref{eq:clasop}) is then multiplicative,
\begin{equation}
\hat{q}\,\psi(q,p)=q\psi(q,p)~,\qquad \hat{p}\,\psi(q,p)=p\psi(q,p)~.
\end{equation}
As is well known from classical statistical mechanics, the evolution of a classical probability distribution $\rho_{\mbox{\tiny C}}(q,p)$ is given by the Liouville equation
$\partial_t \rho_{\mbox{\tiny C}} = \hat L \rho_{\mbox{\tiny C}}$~\cite{Ballentinebook}.  Since the Liouville operator $\hat{L}$ is linear in the derivatives, the evolution equation for $\psi$ is the same as that for $\rho_{\mbox{\tiny C}}$:
\begin{align}
i\hbar \partial_t \psi=&{} \hat{H}_{\mbox{\tiny CL}} \psi~,
\qquad \hat H_{\mbox{\tiny CL}}:=  i\hbar \hat L= \widehat{\partial_p H_{\mbox{\tiny C}}}~\hat{p}_q + \widehat{\partial_q H_{\mbox{\tiny C}}}~\hat{q}_p~.
\end{align}
Here $H_{\mbox{\tiny C}}(q,p)$ is the classical Hamiltonian and we have defined the operators $(\hat p_q,\hat q_p)$ to be  canonically conjugate  to   $(\hat{q},\hat{p})$:
\begin{equation}
[\hat{q},\hat{p}_q]=i\hbar=[\hat{q}_p,\hat{p}]~,\qquad [\hat{q},\hat{q}_p]=[\hat{p},\hat{p}_q]=0~;
\end{equation}
which are correspondingly represented by
\begin{equation}
\hat{p}_q=-i\hbar \partial_q~,\qquad\hat{q}_p=i\hbar \partial_p~.
\end{equation}
In the following these variables will be called \emph{unobservable} variables.

Moreover, under reasonable assumptions about the  classical  Hamiltonian, the Liouville Hamiltonian $\hat{H}_{\mbox{\tiny CL}}$ is   essentially self-adjoint   in the inner product,
\begin{equation}
(\psi,\phi)=\int \text{d}q\text{d}p\,\psi^*\phi~,\qquad\psi,\phi\in\overline{\mathcal{H}}_{\mbox{\tiny C}}~,
\label{eq:inner}
\end{equation}
so it generates a unitary evolution~\cite{ReedSimon1981}. The completion of $\overline{\mathcal{H}}_{\mbox{\tiny C}}$ in the inner product (\ref{eq:inner}) constitutes the classical Hilbert space $\mathcal{H}_{\mbox{\tiny C}}$.

Hereafter let us work for convenience in the Heisenberg picture in which the operators (\ref{eq:clasop}) carry the time dependence. With the previous definitions it is direct to check that the Heisenberg equations for the operators $(\hat{q},\hat{p})$ have the same form as the classical Hamilton equations. The unobservable variables $(\hat{q}_p,\hat{p}_q)$ have their own evolution without influencing the physical sector $(\hat{q},\hat{p})$. So, at least at this level, their evolution is irrelevant. Let us point out that, concerning the classical observable sector and directly related to this decoupling, there exists a symmetry under phase  transformations of the form
\begin{equation}
\psi(q,p)\longrightarrow e^{i\varphi(q,p)}\psi(q,p)~.
\label{eq:gauge}
\end{equation}
As we will see, this ambiguity disappears when developing a CQ interaction scheme. To end this summary, let us mention that it can be rigorously proved~\cite{Koopman1931,Neumann1932} that the Koopman-von Neumann-Sudarshan formulation of classical mechanics is equivalent to the standard one in terms of symplectic manifolds and permits the use of operator techniques to treat classical problems. In particular, it is especially useful to study some aspects of statistical mechanics and ergodic theory~\cite{ReedSimon1981}.

Let us now consider an interacting CQ system~\cite{Sudarshan1976,PeresTerno2001}. The space of states is the tensor product $\mathcal{H}_{\mbox{\tiny C}} \otimes\mathcal{H}_{\mbox{\tiny Q}}$, and the total Hamiltonian operator is
\begin{equation}
\hat{H}_{\mbox{\tiny T}}=\hat{H}_{\mbox{\tiny CL}}+\hat{H}_{\mbox{\tiny Q}}+
\hat{H}_{\mbox{\tiny I}}~,
\end{equation}
where $\hat{H}_{\mbox{\tiny Q}}$ stands for the Hamiltonian of the quantum system and $\hat{H}_{\mbox{\tiny I}}$ provides the CQ interaction. At this point, the first question with a nonstraightforward answer appears: How do we determine the form of the operator $\hat{H}_{\mbox{\tiny I}}$ from its classical counterpart $H_{\mbox{\tiny I}}$?

To discuss this point and connect with earlier work on the subject, let us first analyze  a simple example, which was the object of study in~\cite{PeresTerno2001}:  two harmonic oscillators, one classical and one quantum, 
\begin{equation}
\hat{H}_{\mbox{\tiny CL}}= \hat{p} \hat{p}_q + \Omega^2 \hat{q} \hat{q}_p,~~~
\hat{H}_{\mbox{\tiny Q}}= {1 \over 2}\left(\hat{k}^2 +\omega^2\hat{x}^2\right),
\end{equation}
with bilinear classical coupling
\begin{equation}
H_{\mbox{\tiny I}}=\gamma qx~.
\label{eq:clasint}
\end{equation}
Given any interaction term, one can easily check that the classical Hamiltonian equations of motion (considering the system completely classical) and quantum Heisenberg equations are formally equivalent. The proposal of Peres and Terno~\cite{PeresTerno2001} as the \emph{definite benchmark} for an acceptable classical-quantum hybrid was that any hybridization of this system should respect this equivalence. Then, the result which they proved is that even for this simple system there is no term $\hat{H}_{\mbox{\tiny I}}=\hat{H}_{\mbox{\tiny I}}(\hat{q},\hat{p},\hat{x},\hat{k},\hat{q}_p,\hat{p}_q)$ which fulfills this condition, so they rejected this kind of hybrid CQ dynamics for not being physically meaningful.

In brief, the problem emerges as follows. If one wants the operators of the quantum sector $(\hat{x},\hat{k})$ to appear in the equations of motion of the classical variables $(\hat{q},\hat{p})$, one has to introduce the unobservable variables $(\hat{q}_p,\hat{p}_q)$ in the interaction term. But, by doing that, the decoupling of the unobservable sector no longer holds. That is, the equations of motion of the variables in the physical sector will contain the unobservable operators explicitly. As the unobservable operators do not appear in the QQ theory, the equations of motion of the CQ and QQ theories are different.

With the risk of overinterpreting Peres and Terno's logic, in our view they took this condition as a \emph{definite benchmark} because they identified formally having  equal Heisenberg equations with obtaining an appropriate correspondence principle. As we will see, these are, in principle, logically distinct issues. Hereafter, we will call the Peres-Terno \emph{benchmark} the condition of just obtaining the same correspondence limit starting from the QQ system and from the CQ hybrid.

As mentioned in the Introduction, this benchmark is fully appropriate if one assumes that the hybrid theory is just a particular approximation of the straight QQ theory (here by straight we mean the standard quantization one would have performed to the classical Hamiltonian for two interacting harmonic oscillators). However, if one is looking at hybrid systems as examples of new physics, then the violation of this benchmark could be interpreted positively. The precise prescription of an interaction term $\hat{H}_{\mbox{\tiny I}}$ should take into account additional physical insights coming from the detailed characteristics of the variable that is being regarded as classical.    

Having this in the back of our minds, let us, however, proceed in this section to see whether it is really necessary to abandon the Peres-Terno benchmark. First, we will discuss the simple quadratic interaction term~(\ref{eq:clasint}), and then we will comment about more complicated interactions.

The starting point of our discussion is the observation that the mixing of the unobservable and physical sectors produces an enlargement of the total relevant degrees of freedom of the system. On these grounds the unobservable sector should be constrained in some way to allow the comparison of  the hybrid theory and the CC and QQ theories  on equal footing. To do that, we are going to describe the different CQ, CC, and QQ systems with the (infinite) collection of evolution equations associated with the hierarchy of moments of the corresponding distribution functions of the systems~(see~\cite{Ballentine1998} for a treatment of the CC and QQ cases). The advantage of this formulation is that it permits a step-by-step study of the relevant constraints at different orders.

Among all the interaction terms which can be defined let us analyze
\begin{equation}
\hat{H}_{\mbox{\tiny I}}=\gamma\left(\frac{\hat{q}}{2}+\hat{q}_p\right)\hat{x}~,
\label{eq:iterm}
\end{equation}
which, as we will see, has interesting properties. This interaction leads to the equations of motion:
\begin{align}
\dot{\hat{q}}&=\hat{p}~, & \dot{\hat{p}}&=-\Omega^2\hat{q}-\gamma\hat{x}~,
\nonumber\\
\dot{\hat{x}}&=\hat{k}~, & \dot{\hat{k}}&=-\omega^2\hat{x}-\gamma\left(\frac{\hat{q}}{2}+\hat{q}_p\right)~,
\label{eq:hei2}\\
\dot{\hat{q}}_p&=\hat{p}_q~, & \dot{\hat{p}}_q&=-\Omega^2\hat{q}_p-\frac{\gamma}{2}\hat{x}~.
\nonumber
\end{align}

The equations for the first  moments can be read directly from these formulas: One just has to take the mean values on both sides of the equations.  
Now, it is easy to see that if one imposes the conditions
\begin{equation}
\langle\hat{q}_p+\frac{1}{2} \hat{q}\rangle=\langle\hat{q}\rangle~,\qquad \langle\hat{p}_q+\frac{1}{2} \hat{p}\rangle=\langle\hat{p}\rangle~,
\label{eq:firstcons}
\end{equation}
which form a closed set under time evolution (that is, they are fully consistent as a set of constraints), it is possible to actually satisfy the benchmark. In fact, we have checked that the interaction term (\ref{eq:iterm}) is the only term involving only position operators, $\alpha\hat{q}\hat{x}+\beta\hat{q}_p\hat{x}$, with $\alpha$ and $\beta$ being real parameters, which permits the recovery of the benchmark through the imposition of consistent constraints. This precise term can be obtained by applying the general CQ formulation developed in~\cite{Jauslin2011}, a formulation which suffers from the same problems pointed out by Peres-Terno, to this particular system. An additional good property of the interaction term (\ref{eq:iterm}) is that, as we have checked, it does not lead to strange behaviors such as the energy nonconservation which appears in the particular model analyzed by Peres and Terno. 
Let us also mention in passing that this interaction term does not belong to the class analyzed by Sudarshan; that is, it does not satisfy the classical integrity criterion~\cite{Sudarshan1976,Sherry1979}.

To gain insight into the significance of the appearance of the up-to-now unobservable variables in the hybrid dynamics, let us come back to the local phase symmetry (\ref{eq:gauge}) of the classical observable sector. The introduction of an interaction breaks this symmetry, making the unobservable variables actively participate  in  the behavior of the physical sector. That is, different elections of the local classical phase result in different evolutions and so in different active degrees of freedom. This can be seen by looking at the classical wave function $
e^{i\varphi(q,p)}\psi(q,p)$.
The quantity $|\psi(q,p)|^2$ contains all the information about the moments of the classical variables $(\hat{q},\hat{p})$. On the other hand,  the moments of the unobservable pair $(\hat{q}_p,\hat{p}_q)$ involve  the correlations between the unobservable and physical sectors, and therefore depend on the phase $\varphi(q,p)$. In particular, a change in this local phase results in different mean values $\langle\hat{q}_p\rangle$, $\langle\hat{p}_q\rangle$, which are the relevant quantities to the benchmark [recall Eq. \ref{eq:firstcons}]. In view of this, the local phase $\varphi(q,p)$~, which is irrelevant when there is no interaction between the classical and quantum sectors, has great influence over the properties of the hybrid system.

Although we have succeeded in fulfilling the benchmark, from the paragraph above we see that the hybrid theory still has many more active degrees of freedom than its corresponding CC and QQ counterparts. To set up a proper comparison between these theories one needs to impose additional constraints in the hybrid theory. This suggests analyzing the behavior of the higher moments in the hybrid theory. The second moments are almost as interesting as the first ones: It is well known that for a general quantum system, the  equations for the moments of a classical system and a quantum system coincide for the first and second moments and may start to differ  for higher moments~\cite{Ballentine1998}. For quadratic systems this equivalence  extends to all moments. Then, since the QQ and CC systems have the same second-moment equations, following Peres-Terno logic, it appears sensible to require that the hybrid CQ system also follows these same equations. We will here call this condition the {\em second-moment benchmark}.

To simplify this analysis we find it convenient to work in a new set of canonical variables:
\begin{align}
 \hat{\bar q}&:= \hat{q}_p + {1 \over 2}\hat{q}~,& \hat{\bar p} &:=\hat{p}_q + {1 \over 2}\hat{p}~,
\\
 \hat{l}_p&:=\hat{q}_p - {1 \over 2}\hat{q}~,& \hat{l}_q&:=\hat{p}_q - {1 \over 2}\hat{p}~.
\end{align}
Then the CQ Hamiltonian is expressed as
\begin{align}
\hat{H}_{\mbox{\tiny T}}&= -{1 \over 2}\left(\hat{l}_q^2
 +\Omega^2\hat{l}_p^2 \right) \nonumber\\
 &+ {1 \over 2} \left(\hat{\bar p}^2 +\Omega^2\hat{\bar q}^2\right) + {1 \over 2}\left(\hat{k}^2 +\omega^2\hat{x}^2\right) + \gamma \hat{\bar q} \hat{x}~,
\end{align}
and the first-order equations of motion become
\begin{align}
\dot{\hat{{\bar q}}}&=\hat{\bar p}~, &\dot{\hat{\bar p}}&=-\Omega^2\hat{\bar q} -\gamma \hat{x}~,
\\
\dot{\hat{x}}&=\hat{k}~,&\dot{\hat{k}}&=-\omega^2\hat{x} -\gamma \hat{\bar q}~,
\\
\dot{\hat{l}}_p&=\hat{l}_q~,&\dot{\hat{l}}_q&=-\Omega^2\hat{l}_p~.
\end{align}

As we explained before, for the first-moment benchmark we need the constraint conditions (\ref{eq:firstcons}), which in these new variables read 
\begin{equation}
\langle  \hat{l}_p\rangle=0~,\qquad\langle  \hat{l}_q\rangle=0~.
\end{equation}
In order to satisfy the second-moment benchmark one needs all the moments involving variables $\bar q$ and $\bar p$ to yield the same results that if one would have occurred $q$ and $p$ were used instead. By performing a recursive analysis of constraints 
it is not difficult to check that one needs to impose the following set of constraints:
\begin{align}
\langle \hat l_p \hat {\bar q}\rangle&=0~, &
\langle \hat l_p \hat {\bar p}\rangle&=0~, &
\langle \hat l_p \hat x\rangle&=0~, &
\langle \hat l_p \hat k\rangle&=0~,
\nonumber\\
\langle \hat l_q \hat {\bar q}\rangle&=0~, &
\langle \hat l_q \hat {\bar p}\rangle&=0~, &
\langle \hat l_q \hat x\rangle&=0~,&
\langle \hat l_q \hat k\rangle&=0~, 
\label{eq:secondconst1}
\\
\langle \hat l_p^2\rangle&=0~,&
\langle \hat l_q^2\rangle&=0~,& &&
\hspace*{-10em}\langle \hat l_p \hat l_q + \hat l_q \hat l_p \rangle&=0~.
\nonumber\end{align}
This set is closed under time evolution so that it is consistent with the dynamics of the system. The same set can be equivalently obtained by working with the original coordinates (restricting to quadratic Hamiltonians, there are no operator ordering problems).

By looking at the previous set of constraints one immediately realizes that they cannot be  simultaneously satisfied. The variables $\hat{l}_q$ and $\hat{l}_p$ are canonically conjugate, so one cannot impose that their quadratic mean values $\langle \hat l_q^2\rangle$ and
$\langle \hat l_p^2\rangle$ be simultaneously zero. Therefore, the second-moment benchmark cannot be attained. This result tells us that, even if the correspondence principle (first-moment benchmark) can be preserved by the definition of a suitable interaction term (\ref{eq:iterm}) and imposing suitable constraints, the second order moments will inevitably follow a different dynamics than both QQ and CC systems. This happens independently of whether or not one introduces some consistent constraints in the definition of the CQ system.

The final situation is the following. If one imposes the second-moment benchmark as a necessary condition for a CQ theory to make sense, then these hybrid systems are unphysical. However, this option follows from the assumption that quantum mechanics is a fundamental theory from which the classical theories emerge.
When trying to explore the validity of this very assumption, it is not sensible to directly reject these types of theories. If one decides to accept these hybrid theories as interesting systems to confront with reality, one still has two options. 

On the one hand, one could decide not to impose any constraint on the Sudarshan CQ system. This would imply that the CQ hybrid would have  intrinsic new degrees of freedom. One would need additional physical information about the system one is trying to describe in order to obtain an interpretation of these new degrees of freedom. 

On the other hand, one might still want to have a CQ hybrid theory with the same number of degrees of freedom as the QQ system (either because a specific physical situation asks for it or because one wants to carry out a more direct or strict comparison between both CQ and  QQ theories). To deal with these situations one would need to introduce some consistent constraints that eliminate the irrelevant number of degrees of freedom. It is interesting to note that a constraint of this type could be, for example,
\begin{equation}
\hat H_{\mbox{\tiny T}}\Psi(x,q,p)=0~.
\end{equation}
Notice that if one interprets the operator 
\begin{equation}
\hat E :={1 \over 2}\left(\hat{l}_q^2 +\Omega^2\hat{l}_p^2 \right)
\end{equation}
as a tipe of energy and represents it as $i\partial_t$, one ends up with a Schr\"odinger system equivalent in terms of degrees of freedom to the QQ system we started from.
As is well known, this effective time is not completely equal to the time appearing in Schr\"odinger's equation due to the bounded character of $\hat E$. However, it can be interpreted as an effective time parameter as is done in quantum cosmology~\cite{Unruh1989}. We think it is interesting to note the formal similarity of these types of constrained theories with those appearing in quantum cosmology, which is precisely the system for which the direct application of pure quantum notions is more controversial.

One way or another, one needs additional information to work with hybrid systems, either in the form of specifying additional initial conditions or by  prescribing how to conveniently restrict the total number of degrees of freedom. 
This information depends crucially on the specific character of the variable to be treated (close to) classically. As a taste of the physics which might be modeled with this formalism, we can think of the interaction term (\ref{eq:clasint}) as some sort of continuous measurement of the quantum variable $\hat{x}$. By looking at the Heisenberg equations of motion (\ref{eq:hei2}), one realizes that the difference between this CQ theory and the QQ theory is located in the evolution equation of the quantum momentum $\hat{k}$. So this formalism might take into account the effect of quantum measurements of the variable $\hat x$ by classical devices in a self-consistent hybrid framework. It is also interesting to note that, from this point of view, one could decide to regard the additional degrees of freedom as representing an environment, similar to the environment variables in the standard approach of quantum decoherence (see, for example,~\cite{Zurek2003}). Then, one could trace over the degrees of freedom of this effective environment. Although we have already noted this in the Introduction, let us stress again that, of course, we do not think of this construction as a full theory of the interaction of classical and quantum systems, but merely as a toy model which, however, could be useful to explore some of the features of this hypothetical CQ theory.

To end this section, let us comment on some aspects of the generalization of this formalism to systems with classical a Hamiltonian of the type 
\begin{equation}
H_\text{\tiny T}=\frac{1}{2}p^2+\frac{1}{2}k^2+V(q,x)~,
\end{equation}
where $V(q,x)$ is a suitable interaction potential that is not necessarily quadratic in its variables. It is well known that, when $V(q,x)$ is not quadratic, then the equations of motion of the different   moments do not decouple~\cite{Ballentinebook,Ballentine1998}. For instance, in the equations of motion of the mean values of the canonical variables the higher-order moments will appear. This obstructs the possibility of recovering the first-moment benchmark by imposing first-moment  constraints. Together with the previously described impossibility of satisfying the second-moment benchmark for quadratic systems, this dynamical mixing strongly suggests that with nonquadratic interactions one would not be able to recover, in general, even the first-moment benchmark.

\section{Statistical consistency of hybrids\label{Sec:Others}}

In this section we want to briefly discuss  the ``statistical consistency problem'' of hybrid formulations raised recently by Salcedo~\cite{Salcedo2012} (in this section we follow his analysis, although slightly changing the presentation). To describe this issue in its simplest terms, let us consider a CQ theory in which the quantum system is described as a classical system with all the information of the quantum state contained in a set of symplectic pairs $(X_i,K_i)$~\cite{Heslot1985,Elze2011}. These pairs are nothing more than the complex coefficients $a_i=X_i +i K_i$ in the expansion of the quantum state in a basis of the Hilbert space $\Psi=\sum_i a_i \psi_i$. The evolution of these variables is dictated by a formally classical Hamiltonian of the form   
\begin{eqnarray}
&&H_{\mbox{\tiny Q}}(X_i,K_j) = \sum_{ij} H_{ij}\left(X_i +i K_i\right)\left(X_j -i K_j\right)~;
\\
&&\hspace{2cm} H_{ij}=\langle \psi_i |\hat H |\psi_j\rangle~.
\end{eqnarray}
Equivalently, the dynamical evolution of the expectation value of any observable can be computed directly by just finding the specific expression of the observable in terms of $X_i,K_i$: 
\begin{eqnarray}
&&\langle \hat A\rangle (X_i,K_j) = \sum_{ij} A_{ij}\left(X_i +i K_i\right)\left(X_j -i K_j\right)~;
\\
&&\hspace{2cm} A_{ij}=\langle \psi_i |\hat A |\psi_j\rangle~.
\end{eqnarray}

In order to deal with mixed states, in standard quantum mechanics one introduces the density matrix $\hat \rho = \sum_{ij} \rho_{ij} |\psi_i \rangle \langle \psi_j|$ such that ${\rm Tr}(\hat \rho)=1$. Knowing the density matrix of the quantum system amounts to knowing the value of the Hermitian, positive, and unit-trace matrix $\rho_{ij}$.  For an $N$-state quantum system, the specification of a pure quantum state requires $2N$ real numbers (one less if one considers that the state is unit norm). The specification of a generic density matrix requires $N^2$ real numbers [one less if one imposes the condition ${\rm Tr}(\hat \rho)=1$].

However, if we were interpreting the quantum system completely in classical terms, when extending the formalism to incorporate statistical ensembles of states, instead of the standard density matrix, we would introduce a positive and normalized function ${\cal P}(X_i,K_j)$. The amount of information encoded in this function is much larger than in a density matrix, but in a purely quantum evolution this extra information does not play any role. 

To see this clearly consider, for example, a spin-$1/2$ system ($N=2$). A density matrix can be written in different representations. For instance,
\begin{align}
\rho&= {1 \over 2}|\uparrow \rangle \langle \uparrow|+
{1 \over 2}|\downarrow \rangle \langle \downarrow|
\nonumber\\
&
={1 \over 2}|\leftarrow \rangle \langle \leftarrow|+
{1 \over 2}|\rightarrow \rangle \langle \rightarrow|~.
\end{align}
Here $|\uparrow \rangle$ and $|\downarrow \rangle$ represent the eigenstates of the spin in the $z$ direction, and $|\leftarrow \rangle$ and $|\rightarrow \rangle$ represent those in the $y$  direction. These are just different representations; the density matrix is one and the same.
However,  the probability functions corresponding to each of these representations are actually different: 
\begin{align}
{1 \over 2} \delta^{4}(\xi_a-\xi_a^\uparrow)
+{1 \over 2} \delta^{4}(\xi_a-\xi_a^\downarrow)&:={\cal P}_1(\xi_a)
\nonumber\\
\neq
{1 \over 2} \delta^{4}(\xi_a-\xi_a^\leftarrow)
+{1 \over 2} \delta^{4}(\xi_a-\xi_a^\rightarrow)&:={\cal P}_2(\xi_a)~,
\end{align}
where $\xi_i =X_i$, $\xi_{2+i} =K_i$, and  $\xi_a^\sigma$ represent the $X_i$ and $K_i$ values associated with the corresponding quantum state $|\sigma \rangle$. 
If the quantum system does not interact with a classical sector, this difference does not lead to observable effects. In fact, two distribution functions ${\cal P}_1(\xi_a)$ and ${\cal P}_2(\xi_a)$ such that
\begin{align}
\int \text{d}\boldsymbol{\xi} &\left[ {\cal P}_1(\xi_a)-{\cal P}_2(\xi_a)\right] 
\nonumber \\ 
&\times\Big[\sum_{ij} A_{ij}\left(X_i +i K_i\right)\left(X_j -i K_j\right)\Big]=0~,
\label{indistinguishability}
\end{align}
for any Hermitian matrix $A_{ij}$, cannot be distinguished. 

However, the issue comes about when adding a generic CQ hybrid interaction,
\begin{eqnarray}
H_{\mbox{\tiny T}} = H_{\mbox{\tiny C}}(q,p) +H_{\mbox{\tiny Q}}(X_i,K_j) + H_{\mbox{\tiny I}}(q,p,X_i,K_j)~.
\end{eqnarray}
Then, the distribution functions will have the form ${\cal P}_{\mbox{\tiny T}}(\xi_a,q,p)$. Imagine that, before the interaction is switched on, we start  with two different distribution functions with equal and separable classical parts: 
\begin{align}
{\cal P}_{\mbox{\tiny T1}}(\xi_a,q,p)&={\cal P}_1(\xi_a)\rho_{\mbox{\tiny C}}(q,p)~,
\nonumber\\
{\cal P}_{\mbox{\tiny T2}}(\xi_a,q,p)&={\cal P}_2(\xi_a)\rho_{\mbox{\tiny C}}(q,p)~,
\end{align}
such that  
${\cal P}_1(\xi_a)$ and ${\cal P}_2(\xi_a)$ initially satisfy~(\ref{indistinguishability}) for any Hermitian matrix $A_{ij}$. Notice that now the coefficients $A_{ij}$ derive from hybrid observables:
\begin{eqnarray}
A_{ij}=\int \text{d}q \text{d}p~\rho_{\mbox{\tiny C}}(q,p) A_{ij}(q,p)~.
\end{eqnarray}
The hybrid dynamics makes it, in general, impossible to express the time derivative of the observable in~(\ref{indistinguishability}) ,
\begin{eqnarray}
&&\sum_{ij} \left[(\partial_q A_{ij}(q,p)) \dot q + (\partial_p A_{ij}(q,p)) \dot p \right]
\left(X_i +i K_i\right)\left(X_j -i K_j\right)
\nonumber \\ 
&&\hspace{0.5cm}
+A_{ij}(q,p)(\dot X_i +i \dot K_i)(\dot X_j -i \dot K_j)~,
\label{derivative}
\end{eqnarray}
as a quadratic form
\begin{eqnarray}
\sum_{ij} B_{ij}(q,p)\left(X_i +i K_i\right)\left(X_j -i K_j\right)~.
\end{eqnarray}
Therefore, the dynamical evolutions of initially equal density matrices separate, and the additional information in the probability functions comes into play.

Here we want to highlight that this phenomenon can be interpreted as emergent new physics caused by hybridization (something already acknowledged by Salcedo but with the opposite emphasis) and that the effect is similar to what we have seen for Sudarshan hybrids. The information contained in a hybrid state is more than that contained in the product of a quantum density matrix and a classical distribution function. 

Consider, for instance, the highly hypothetical case in which the system could be modeled satisfactorily as a pure quantum sector and a pure classical sector without interaction for all times $t<t_0$. If an interaction between the two sectors is switched on at $t_0$,  the formalism tells us that, in order to know how the system will evolve, it will not be  enough to know the classical distribution and the quantum density matrix at the time $t_0^-$. As additional information we would need to select a specific representation of the density matrix as being special. From the point of view of a quantum system, this information will qualify as hidden variables \footnote{An anonymous referee made us notice that there is an analogy to this behavior in de Broglie-Bohm quantum mechanics. To calculate the trajectory of one particle in a two-particle system it is not enough to know the initial position and velocity of the particle and its reduced density matrix, one has to specify the entire wave function of the two-particle system. Different wavefunctions having the same reduced density matrix could lead to completely different particle trajectories.}. The only source of information available that could help select a specific representation is contained in the very form of the interaction terms. However, it might seem that this new information is somewhat arbitrary, leaving the theory with no predictive power.

However, consider now the more realistic case in which the CQ interaction was always present and thus that the state of the system was always hybrid. The state of the system would always have had this additional information. By controlling the interaction one can imagine driving the system to a situation in which part of the system behaves as a pure quantum system. The additional information beyond the quantum density matrix will be there, but will be irrelevant for the evolution of the pure quantum sector during its isolation. The additional information, now hidden, would be perfectly dictated by the initial state of the hybrid system and its precise hybrid dynamics. Once the isolation of the quantum part is removed, the evolution will follow a precise track, with no arbitrariness.

We cannot help but seeing the situation as reminiscent of what happens with decoherence in standard quantum mechanics: The specific way in which the environmental variables interact with the system determines the specific pointer basis in which decoherence occurs (see, e.g.,~\cite{Zurek2003}). In this framework the state of the environment and the form of its interactions play the role of hidden variables.

\section{Two oscillators in the Wigner representation \label{Sec:Wigner}}

In this section, we want to describe some characteristic behavior expected generically in hybrid systems. With this aim we are going to deal with a simple model based on the Wigner quantization of a quadratic system composed of two interacting harmonic oscillators, one classical and one quantum. As we will see, the distinction between the classical and the quantum sectors in this model is just a matter of initial conditions for the distribution functions and has nothing to do with the dynamics, which is equal for both sectors. In this sense, the model does not qualify as a proper hybrid. Let us also mention that this dynamics cannot be recovered within the set of Sudarshan evolutions. The Sudarshan analysis points out the presence of new physics even at the level of simple quadratic models. Having said that, here we will use this model exclusively to illustrate how the initial dispersion in the initially quantum variables is modified by the interaction with the initially classical sector, thus blurring the classical-quantum distinction.

It is well known that quantum mechanics can be alternatively formulated as a phase-space theory, formally equivalent to classical mechanics~\cite{Moyal:1949sk,Groenewold:1946kp}. In classical mechanics, for each function in phase space $A(q,p)$, one associates an operator (a Hamiltonian vector field) 
\begin{equation}
\hat A= \{A,\cdot\}_\text{\tiny P}~,
\end{equation}
where $\{\cdot,\cdot\}_\text{\tiny P}$ represents the Poisson  bracket. Then, one can define the Lie algebra in the space of operators as  
\begin{eqnarray}
[\hat A, \hat B] :=\hat A \cdot \hat B - \hat B \cdot \hat A = \widehat{\{A,B\}}_\text{\tiny P}~.
\end{eqnarray}
Here the dot (~$\cdot$~) denotes the straightforward consecutive application of the operators.
The evolution of any dynamical variable is then dictated by either of the following equivalent equations:
\begin{eqnarray}
{\text{d} \hat A \over \text{d}t}=[\hat A, \hat H]+{\partial \hat A \over \partial t}~,\qquad
{\text{d}A \over \text{d}t}=\{A,H\}_\text{\tiny P} + {\partial A \over \partial t}~.
\end{eqnarray}
Finally, a classical statistical theory can be constructed by defining a probability distribution function $\rho_\text{\tiny C}(q,p)$.

In the Heisenberg picture a particular quantization of a classical system (which involves the selection of a specific operator ordering) can be constructed in the same formal way by simply replacing the Poisson bracket by the Moyal bracket~\cite{Moyal:1949sk}:
\begin{align} 
\{f,g\}_\text{\tiny M}:=&\frac{1}{i\hbar}(f\star g-g\star f)=\{f,g\}_\text{\tiny P}+o(\hbar)~,\\
f\star g:=&f(x,k)\exp\left[{\frac{i\hbar}{2}(\stackrel{\leftarrow }{\partial }_x \stackrel{\rightarrow }{\partial }_{k}-\stackrel{\leftarrow }{\partial }_{k}\stackrel{\rightarrow }{\partial }_{x})} \right] g(x,k)~. 
\end{align}
The classical probability distribution function is now substituted by the Wigner distribution function $W_\text{\tiny Q}(x,k)$.

In this setting the construction of a hybrid theory would involve the construction of a hybrid bracket. As suggested by what is known in the literature, the straightforward construction of such a bracket could be obstructed if one requires certain conditions~\cite{Salcedo2012}. However, one could certainly do it in an extended version, such as the Sudarshan analysis in the previous sections, but at the cost of having many different possible constructions.

The simplicity of the analysis that follows comes from the selection of a quadratic system. When restricted to quadratic systems, Moyal and Poisson brackets coincide, so the search for a hybrid bracket can be seen as trivial: One can choose this unique bracket as a hybrid bracket. Within this Wigner-type approach the quadratic systems do not openly call for new dynamical physics, but still serve to draw our attention to other emergent phenomena. We expect that Sudarshan hybrids (and other hybrid formulations) will exhibit qualitative behaviors similar to the ones described by this toy model. 

We want to study the behavior of a hybrid distribution function $W_\text{\tiny CQ}(q,p,x,k,t)$.  As before, let $(q,p)$ be the phase-space variables that describe one of the oscillators, for instance, the one with classical initial conditions. Similarly, let $(x,k)$ be the position and canonical momentum of the second oscillator, which will have quantum initial conditions. Let $\Omega$ and $\omega $ be their corresponding characteristic frequencies. Finally, let us consider that these oscillators are coupled by an interaction term of the form $\gamma q x$, as before. Then the  classical Hamiltonian of this system is obviously
\begin{equation}
H_\text{\tiny T}=\frac12(p^2+\Omega^2q^2)+\frac12(k^2+\omega ^2x^2)+\gamma qx~.
\end{equation}
The classical trajectories can then be solved to yield
\begin{align}
	\mathbf{\boldsymbol \xi}(t)= U(t)\mathbf{\boldsymbol \xi}'~,
    \label{eq:evolxi}
\end{align}
where $\mathbf{\boldsymbol \xi}$ is the column vector $\mathbf{\boldsymbol \xi}=(q,p,x,k)^\textsc{t}$, $\mathbf{\boldsymbol \xi}'$ are the initial conditions, and the evolution matrix $U(t)$ can be obtained via a standard expansion in normal modes.

Now, hybrid states of this system of two coupled harmonic oscillators are determined by the hybrid Wigner distribution $W_\text{\tiny CQ}(\mathbf{\boldsymbol \xi},t)$. As we have already mentioned, because for a quadratic system the evolutiona of a classical distribution function (Poisson bracket) and a quantum distribution function (Moyal bracket) coincide, we will use this very same dynamics for the evolution of the hybrid distribution function
\begin{equation}
\partial_t W_\text{\tiny CQ}=\{H_{\mbox{\tiny T}},W_\text{\tiny CQ}\}_\text{\tiny P}~.
\end{equation}
Hereafter, to simplify notation we eliminate the subscript CQ in the complete distribution function.  This equation can be easily solved by means of the time-dependent canonical transformation generated by the inverse classical evolution, which takes any phase-space configuration to its initial conditions; namely, we use as canonical variables 
$\mathbf{\boldsymbol \xi}'=U^{-1}(t)\mathbf{\boldsymbol \xi}$. The new Hamiltonian vanishes identically, and therefore, in these variables the Wigner function $W'(\mathbf{\boldsymbol \xi}',t)$  does not evolve:  $W'(\mathbf{\boldsymbol \xi}',t)=W_0(\mathbf{\boldsymbol \xi}')$. Therefore, if we go back  to the original variables we obtain the Wigner function straightforwardly in terms of the initial Wigner distribution $W_0$:
\begin{equation}
W(\mathbf{\boldsymbol \xi},t)=W_0[\mathbf{\boldsymbol \xi}'(\mathbf{\boldsymbol \xi},t)]=W_0[U^{-1}(t)\mathbf{\boldsymbol \xi}]~.
\end{equation}

Let us compute the evolution of the first and second moments. Let us start with the mean values $\langle\mathbf{\boldsymbol \xi}\rangle$:
\begin{align}
\langle\mathbf{\boldsymbol \xi}\rangle&=\int\text{d}^4\xi\,\mathbf{\boldsymbol \xi}\, W(\mathbf{\boldsymbol \xi},t)=\int\text d^4\xi\,\mathbf{\boldsymbol \xi}\, W_0(U^{-1}\mathbf{\boldsymbol \xi})
\nonumber\\
&=\int\text d^4\xi'\,U\mathbf{\boldsymbol \xi}'\, W_0(\mathbf{\boldsymbol \xi}')=U\langle\mathbf{\boldsymbol \xi}\rangle_0~,
\end{align}
where we have used the fact that $\det U=1$.
As should happen for quadratic Hamiltonians, the mean values obey the classical evolution law, regardless of whether the system is classical or quantum.
As for the second moments,
\begin{align}
\langle\mathbf{\boldsymbol \xi}\,\mathbf{\boldsymbol \xi}^\textsc{t}\rangle&=\int\text{d}^4\xi\,\mathbf{\boldsymbol \xi}\,\mathbf{\boldsymbol \xi}^\textsc{t}\, W(\mathbf{\boldsymbol \xi},t)=\int\text d^4\xi\,\mathbf{\boldsymbol \xi}\,\mathbf{\boldsymbol \xi}^\textsc{t}\,W_0(U^{-1}\mathbf{\boldsymbol \xi})
\nonumber\\
&=\int\text d^4\xi'\, U\mathbf{\boldsymbol \xi}'\,\mathbf{\boldsymbol \xi}'^{\textsc{t}}\,U^\textsc{t} W_0(\mathbf{\boldsymbol \xi}')=U\langle\mathbf{\boldsymbol \xi}\,\mathbf{\boldsymbol \xi}^\textsc{t}\,\rangle_0U^\textsc{t}~,
\end{align}
so that the covariance matrix $\Sigma=\langle\mathbf{\boldsymbol \xi}\,\mathbf{\boldsymbol \xi}^\textsc{t}\rangle-\langle\mathbf{\boldsymbol \xi}\rangle\langle\mathbf{\boldsymbol \xi}\rangle^\textsc{t}$
evolves according to 
\begin{equation}
\Sigma=U\Sigma_0U^\textsc{t}~.
\label{eq:evolcovmat}
\end{equation}
It also obeys the classical evolution, again independent of the classical or quantum characteristics of the  variables.

\begin{figure}
	\begin{center}
		\includegraphics[width=\columnwidth]{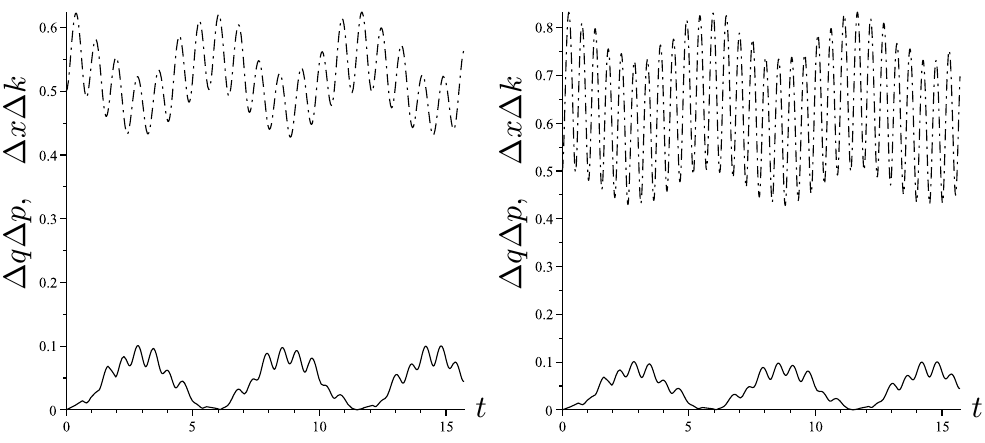}					
	\end{center}
	\caption{In both plots, the solid line shows the time evolution of $\Delta q \Delta p$, and the dashed  one shows $\Delta x \Delta k$. On the left, the frequency of the initially classical oscillator is larger than the frequency of the classical one: $\Omega = 3$ and $\omega =2$. On the right, the situation is reversed, i.e., $\Omega=2$ and $\omega =3$. The coupling has been set to $\gamma=1$.  We also use $\hbar=1$.} \label{fig:f-a}
\end{figure}

\begin{figure}
	\begin{center}
		\includegraphics[width=\columnwidth]{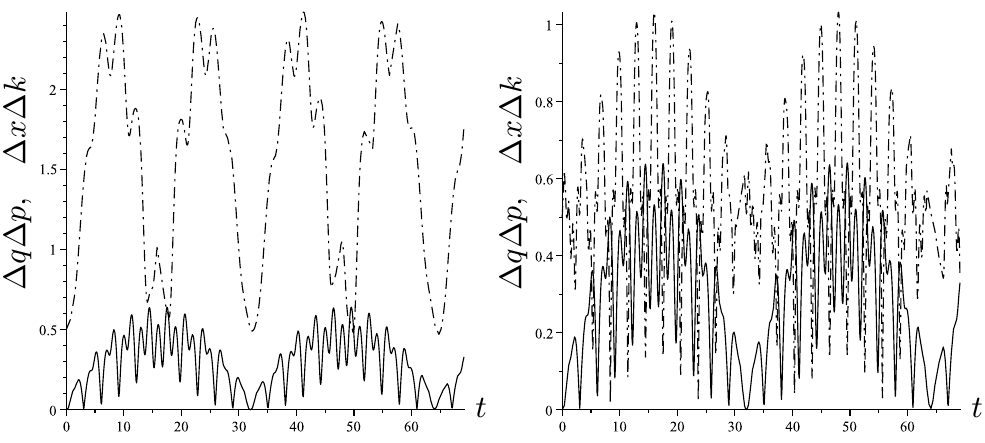}					
	\end{center}
	\caption{In both plots, the solid line shows the time evolution of $\Delta q \Delta p$, and the dashed  one shows $\Delta x \Delta k$. On the left, the  frequency of the initially classical oscillator $\Omega = 2$ is greater than the coupling $\sqrt\gamma=1$, which, in turn, is larger than the frequency of the initially quantum oscillator $\omega =0.51$. 
		On the right,	the situation is reversed, i.e. $\omega  = 2$ and $\Omega=0.51$, with $\sqrt\gamma=1$.  We also use $\hbar=1$.}
		\label{fig:f-b}
	\end{figure}

\begin{figure}
	\begin{center}
		\includegraphics[width=\columnwidth]{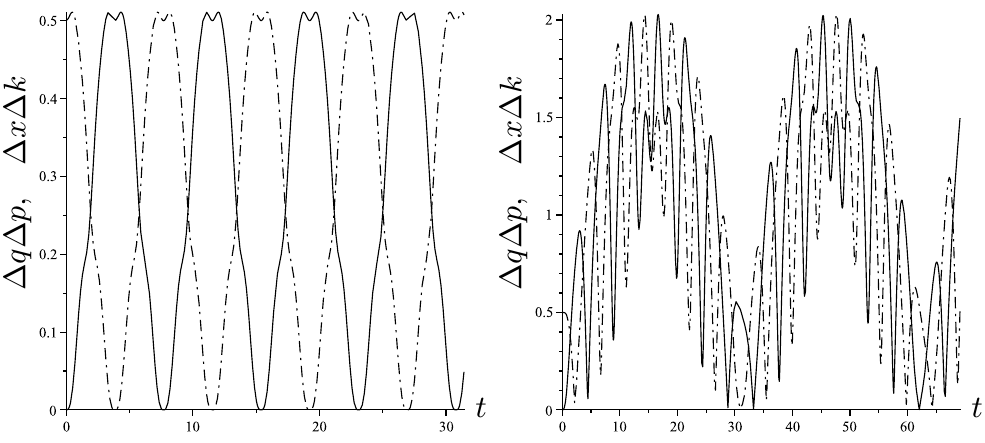}					
	\end{center}
	\caption{In both plots, the solid line shows the time evolution of $\Delta q \Delta p$, and the dashed  one shows $\Delta x \Delta k$. On the left, both frequencies are equal  $\omega _q=\omega _x = 1.73$ and larger than the coupling $\sqrt\gamma=1$. On the right, both frequencies are very similar to each other, with $\omega  = 1.01$,  $\Omega=1$, and to the coupling $\sqrt\gamma=1$. We also use $\hbar=1$.}\label{fig:f-c}
\end{figure}

The classical or quantum nature of the  variables  is entirely encoded in the initial Wigner distribution since, as we have already mentioned, the dynamics is the same in both situations. Our interest is to analyze the dynamics of an initially classical oscillator coupled to an initially quantum one. For the initially classical oscillator, described by the variables $(q,p)$, we will choose a specific initial position and momentum, so that its initial Wigner function will be
\begin{equation}
\rho_{\text{\tiny C}0}(q,p)=\delta(q-q_0)\delta(p-p_0)~.
\end{equation}
The initially quantum oscillator, described by $(x,k)$, will be taken to be initially in a coherent state of the form
\begin{equation}
W_{\text{\tiny Q}0}(x,k)= \frac{1}{2 \pi \sigma_x \sigma_k}e^{-(x - x_0)^2/2 \sigma_x^2}e^{-(k - k_0)^2/2 \sigma_k^2}~.
\end{equation}
Therefore the initial Wigner function of the coupled system will be
\begin{equation}
W_0(q,p,x,k)=\rho_{\text{\tiny C}0}(q,p)W_{\text{\tiny Q}0}(x,k)~,
\end{equation}
so that its first momenta  are
\begin{equation}
\langle\mathbf{\boldsymbol \xi}\rangle_0=\mathbf{\boldsymbol \xi}_0~,
\end{equation}
and the only non-vanishing components of the covariance matrix are
\begin{equation}
(\Delta x)_0^2=\Sigma_{0xx}=\sigma_x^2~,
\qquad
(\Delta k)_0^2=\Sigma_{0kk}=\sigma_k^2~.
\end{equation}
The time evolution of the covariance matrix can be obtained by introducing this result into Eq. (\ref{eq:evolcovmat}). We are now ready to analyze the behavior of the uncertainties of each of the oscillators, given by the diagonal elements of this covariance matrix. Figures \ref{fig:f-a}-\ref{fig:f-c} show the time evolution of $\Delta q\Delta p$ and $\Delta x\Delta k$ for various configurations determined by the characteristic frequencies of both oscillators, $\Omega$ and $\omega $, and the coupling~$\gamma$.

We have seen that if one (or both) frequency is much larger than the coupling, both oscillators effectively decouple.
As we depart from this limit, the product of the uncertainties of both oscillators $\Delta q\Delta p$ and $\Delta x\Delta k$ starts to oscillate (see Fig. \ref{fig:f-a}); i.e., there is a small transfer of quantum uncertainty back and forth. The case in which only one of the frequencies is larger than the coupling leads to a significantly higher transfer of quantum uncertainties between both oscillators (see Fig. \ref{fig:f-b}). The initially classical oscillator evolves to a state which is  not purely classical, and the initially quantum oscillator evolves to a state which is not purely quantum in a (quasi-)periodic evolution.

The other (more interesting) limit is 
when both frequencies $\Omega$ and $\omega $ are almost equal. In this case, 
for very large frequencies compared with the coupling, we end up in the previous situation. However, for finite frequencies, the exchange of quantum uncertainty between both oscillators may be significant and even 
extreme, to the extent that the initially classical oscillator may become purely quantum and vice versa. This is shown in Fig.~\ref{fig:f-c}.

The masses of the two oscillators are encoded in the Hamiltonian as the inverse of the frequencies, and since classical masses are typically large (macroscopic) and quantum masses are in the nanoworld, we could say, in an attempt to extract consequences for real systems from this oversimplified toy model, that it is impossible to see a hybrid system with exotic features in our regular macroscopic world or in the experiments of the quantum world because of the effective decoupling between them. We would need to go to  specific mesoscopic environments to witness a hybrid system with neither classical nor quantum behavior. This suggests that we should not require a hybrid system to have a completely classical or quantum limit because when they interact there is new physics that we will have to cope with.

An interesting observation is that, independent of the parameters of the hybrid system, the total quantum uncertainty never goes below that initially present. More specifically, the quantity $\Delta q\Delta p+\Delta x\Delta k$ is never smaller than its initial value $\hbar/2$. However, there is no restriction concerning each term individually. In  particular, they can vanish (at different times), as seen, for instance, in Fig. \ref{fig:f-c}. Further, we have started with a vanishing value for the uncertainty of the classical oscillator $(q,p)$, and this has led to no inconsistencies.  The idea that the interaction between classical and quantum variables will lead to a transference of quantumness from the initially quantum sector to the initially classical one and vice versa (or of classicality 
in the other direction) was already pointed out by DeWitt in the 1960s~\cite{Dewitt1962,Dewitt1962b} (see also the comments by Unruh in Ref.~\cite{Unruh1984}). In these references, they argued that this transference makes  the coupling of purely classical systems  and purely quantum systems the only consistent one. However, this analysis suggests that hybrid systems might simply be transferring 
around a certain degree of quantumness between its different parts, never becoming purely quantum or classical. We ask ourselves whether this could not even be a crucial ingredient for evolution in nature.

\section{Summary and conclusions \label{Sec:Summary}}

In this work we have tried to find some generic features of hybrid classical-quantum systems by revising some of the models proposed in the literature.

In Sec.~\ref{Sec:Koopman} we analyzed in some detail the hybrid model proposed by Peres and Terno~\cite{PeresTerno2001}. Their model is presented within the Koopman-von Neumann-Sudarshan formulation of classical mechanics and consists of two harmonic oscillators, one quantum and one classical, in interaction. We find that in the case of a quadratic interaction, as opposed to what was suggested by Peres and Terno, it is indeed possible to recover the correspondence principle starting from the CQ system. For that, one needs to impose some constraints on the possible states, which, on the other hand, seems more than reasonable, given the enlargement of degrees of freedom exhibited by the Koopman-von Neumann-Sudarshan formulation. However, this correspondence principle cannot be extended to the second   moments of the variables as   could have been expected. At second order, this hybrid theory shows new physics (a new dynamical behavior) which cannot be recovered within the complete quantization of the classical model. When the interactions are nonquadratic we see that, generically, it is not possible to find a consistent set of constraints leading to the correspondence principle. So we can conclude that these hybrid theories are not limits of the straight quantum theory: They incorporate new physics. In particular, they contain  new active degrees of freedom which one cannot fix without additional physical information. This new information is related to a more detailed description of the quasiclassical variables, variables of a macroscopic (typically complex) nature.

In Sec.~\ref{Sec:Others} we reviewed the statistical consistency problem recently raised by Salcedo~\cite{Salcedo2012} as an obstruction to obtain sensible hybrid systems at least of a particular type. We show that, again, these hybrid systems exhibit properties beyond what would be obtained from a complete and straight quantization of the system. They are not a partial limit of this completely quantum theory. Among other things, one finds again that these hybrid systems have more degrees of freedom than expected.    

Finally, in Sec.~\ref{Sec:Wigner} we have analyzed a very simple system of two harmonic oscillators, initially, one classical and one quantum, coupled through a quadratic interaction term by using a Wigner-distribution formalism. In this model the classical and quantum character of the different variables is not encoded in their dynamics but only in the type of initial conditions  imposed on them. Within this model we have shown that, depending on the specific values of the frequencies of the classical and quantum oscillators, the products of their respective dispersions,   $\Delta q \Delta p$ and $\Delta x \Delta k$, oscillate between different minimum and maximum values. As initial conditions we have  fixed the classical variables to have zero dispersion; the initial quantum state has been set to be a coherent (minimal quantum uncertainty) state. The hybrid evolution requires that initial classical variables develop dispersion up to some levels. In some situations, when the two frequencies are comparable, the behavior of the initially classical and quantum variables is such that one cannot distinguish them after some time. The product of quantum dispersions $\Delta x \Delta k $ reaches almost zero quasiperiodically (this depends on the commensurability of the frequencies), and the product of the classical dispersions, $\Delta q \Delta p$, reaches $\hbar/2$. However, in all situations we have checked that $\Delta q \Delta p + \Delta x \Delta k \geq \hbar/2$. We interpret this fact as indicating that the total quantumness (or classicality) of the total system is being maintained.

All in all, it appears that one can formulate consistent hybrid theories, but these theories incorporate new physics. They are not suitable limits of a direct quantization of the entire system. We should distinguish this straight quantum theory from the possibility of using a quantum formalism to describe all the physics of the hybrid system, as happens in the Koopman-von Neumann-Sudarshan formalism. Among other things, the new physics appears to bring about new degrees of freedom, which should be associated with the additional level of description of the quasiclassical (macroscopic) variables one would need to prescribe the precise form of the quantum back reaction to these variables. In this respect, the measurement process as presented by the Copenhagen interpretation could be seen as an extreme case of interaction between a classical and a quantum system in which there is no transfer of fluctuations into the classical variables. The measurement theory of Sudarshan goes one step further, allowing some transfer of fluctuations while they do not compromise the classical integrity of the variables. Another interesting lesson is that in a hybrid context pure notions such as quantum or classical are only limiting notions, which are very useful in isolated situations but are, strictly speaking, nonexistent; in a hybrid setting everything might have a certain degree of quantumness (classicality).
 
An issue that has not been dealt with in this paper is the role of relativity in hybridization. Many of the aspects analyzed here are quite independent of whether the hybrid system is relativistic or not. However, one should not forget that relativity entails new conceptual approaches to, e.g., causality or localization. They may well affect the dynamical behavior of hybrid systems and deserve further study. 

At the end of the day, only new experiments in the mesoscopic realm can help us decide whether in the behavior of a complete system there is something beyond quantum dynamics of microscopic quantum constituents. Proposals such that those of Di\'osi and Penrose~\cite{Diosi1987,Penrose1996,Marshalletal2003} or Guirardi \emph{et al.}~\cite{Ghirardietal1985} of new physics at the classical-quantum cut find additional justification in light of these hybrid models. In particular, the proposal that gravity, with its stubborn resistance to quantization, might play a role in going beyond standard quantum mechanics is a powerful idea that we should not forsake.

\acknowledgments

Special thanks go to Lorenzo Luis Salcedo and an anonymous referee for enlightening comments and bringing important references to our attention. The authors also want to thank V\'ictor Aldaya, Luis C. Barbado, Julio Guerrero, Gil Jannes and Francisco L\'opez-Ruiz for helpful discussions. Financial support was provided by the Spanish MINECO through projects  FIS2011-30145-C03-01 and  FIS2011-30145-C03-02, by the Consolider-Ingenio
2010 Program CPAN (CSD2007-00042), and by the Junta de Andaluc\'{\i}a through project FQM219. R.C-R. acknowledge support from CSIC through the JAE predoc program, cofunded by FSE. 


\bibliography{hybrid-varxiv}	
\end{document}